\documentclass[%
 aip,
rsi,%
 amsmath,amssymb,
 reprint,%
]{revtex4-1}

\usepackage{graphicx}% Include figure files
\usepackage{dcolumn}% Align table columns on decimal point
\usepackage{bm}% bold math
\begin{document}

\title{Nonlinear photonic dynamical systems for unconventional computing}

\author{D. Brunner}
\affiliation{D\'{e}partement d'Optique P. M. Duffieux, Institut FEMTO-ST,  Universit\'e Bourgogne-Franche-Comt\'e CNRS UMR 6174, Besan\c{c}on, France.}%

\author{L. Larger}
\affiliation{D\'{e}partement d'Optique P. M. Duffieux, Institut FEMTO-ST,  Universit\'e Bourgogne-Franche-Comt\'e CNRS UMR 6174, Besan\c{c}on, France.}%

\author{Miguel C. Soriano}
\email{miguel@ifisc.uib-csic.es}
\affiliation{Instituto de Física Interdisciplinar y Sistemas Complejos (IFISC, UIB-CSIC), Campus Universitat de les Illes Balears E-07122, Palma de Mallorca, Spain.}%

\date{\today}% It is always \today, today,

\begin{abstract}

Driven by the remarkable breakthroughs during the past decade, photonics neural networks have experienced a revival.
Here, we provide a general overview of progress over the past decade, and sketch a roadmap of important future developments.
We focus on photonic implementations of the reservoir computing machine learning paradigm, which offers a conceptually simple approach that is amenable to hardware implementations.
In particular, we provide an overview of photonic reservoir computing implemented via either spatio temporal or delay dynamical systems.
Going beyond reservoir computing, we discuss recent advances and future challenges of photonic implementations of deep neural networks, of the quest for learning methods that are hardware-friendly as well as realizing autonomous photonic neural networks, i.e. with minimal digital electronic auxiliary hardware.

\end{abstract}

\maketitle

\section{\label{sec:Intro} Introduction}

The complementarity between photonics and electronics has long been recognized \cite{Lohmann1990}.
Today, photonics is mostly used for communication over distances as short as meters and continues moving towards taking over intra- or on-chip communication.
Such computing processors heavily leveraging photonic communication is particularly appealing for neural networks (NN), which fundamentally differ in their architecture from classical computing approaches.
A NN's countless connections between its neurons makes photonics a highly attractive alternative to realizing next generation NN processors \cite{Psaltis1985}.
Most importantly, the number of connections in a NN usually scales quadratic with the number of neurons, and as such the relevance of an energy efficient approach to intra-neuron communication will continue to increase in relevance.

A promising approach is so-called photonic in-memory computing \cite{Rios2015}, where the weight of a connection is encoded in physical properties such as attenuation or phase-shift.
When such connections are optically realized, then a connection's strength is physically encoded simply during the photon's propagation.
This process therefore is entirely passive and promises maximal energy efficiency.
However, the required amount of neurons and hence connections in a full-scale deep NN makes such full implementations a daunting task for NN's of competitive size.
After early work on photonics NNs \cite{Psaltis1985}, they gained significant attention and led to impressive results such as realtime object classification \cite{Li1993}.
However, the electronic computer quickly took over as photonic technology of the day could not keep up with the increasing architectural complexity of more advanced NN concepts.

Around a decade ago photonic NNs experienced a revitalization of interest, which was mostly related to the significant reduction in complexity consequence of the reservoir computing (RC) concept \cite{vandoorne2008toward,RCbook}. The proliferation of neuro-inspired computing notions and the advances in photonic integration technologies are currently placing photonics as a realistic and highly attractive complementation to the electronic computing hegemony \cite{shastri2021photonics}. Here, we provide a general overview of the recent progress in photonic RC, and sketch a roadmap of important future developments in the context of photonic neural networks.

\section{\label{sec:RC} Reservoir Computing}

\begin{figure*}[thb]
	\centering
	\includegraphics[width=0.7\columnwidth]{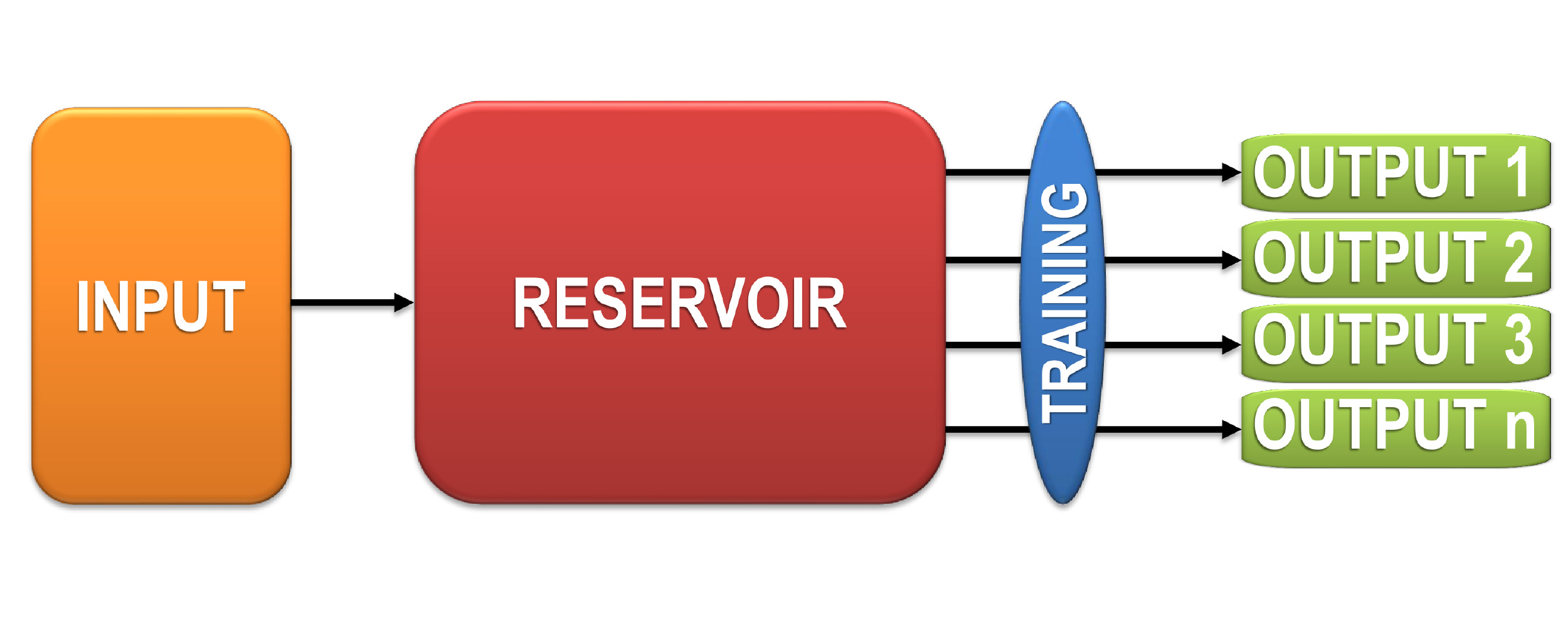}
	\caption{The basic structure for RC is formed by an input layer, a reservoir, and an output layer. The input information is sent to the reservoir, whose internal connections act like a fixed hidden layer in artificial neural networks. The response of the reservoir is used to produce the desired output after an optimization procedure during the training process (only the output connections are adapted in the training). Different sets of output connection weights can be trained in parallel for different tasks. Figure courtesy of Pere Mujal.}
	\label{fig:RC}
\end{figure*}

The revival of photonic neural networks in the last decade has been partly triggered by the appearance of the RC paradigm.
RC is a machine learning technique that was initially developed to provide a simple supervised learning procedure for recurrent neural networks \cite{luko09}. 
Generally speaking, the core of RC is the so-called reservoir, which can be simply seen as a high-dimensional nonlinear dynamical system.
Figure \ref{fig:RC} depicts an illustration of the RC concept with three layers, namely the input layer, the reservoir and the output layer.
Since RC is suitable to solve computationally-hard problems whose solution depend on a temporal context, it has been successfully applied to deal with tasks such as time series prediction or voice recognition \cite{jaeger2004harnessing,verstraeten07}.

A principle quality of RC is that the reservoir can be initialized at random and left untrained.
The learning exclusively focuses on the optimization of an output layer, which is often a mere linear combination of the reservoir's state.
The rather soft requirements for the properties of the reservoir as well as the simplicity of the learning method makes RC highly attractive for photonic hardware implementations \cite{vandoorne2008toward,van2017advances}.
We cover in Sections \ref{sec:spatio} and \ref{sec:delay} two families of RC implementations that have gained attraction on the photonics community, namely photonic RC based on either spatio-temporal or delay-based dynamical systems. 

We reproduce here a relatively simple mathematical formulation of the RC concept. We choose to present the formulation in terms of a popular variant of RC, namely echo state networks \cite{jaeger2002adaptive}:
%\textcolor{red}{HERE WRITE 3-4 lines about the mathematical concept} OK; I can do that.
\begin{eqnarray}
	\mathbf{x}(t+1) = f( \mathbf{W}^{\mathrm{int}}\mathbf{x}(t) + \mathbf{W}^{\mathrm{inj}}\mathbf{u}(t+1) + \mathbf{b} ), \label{eq:RC_state} \\
	\mathbf{y}(t+1) = \mathbf{W}^{\mathrm{out}}\mathbf{x}(t+1), \label{eq:RC_output}
\end{eqnarray}
where we have used a matrix notation for convenience. In Eq. (\ref{eq:RC_state}), $\mathbf{x}(t)$ stands for the state of the reservoir at time $t$, $\mathbf{u}(t)$ is the input information, $f(\cdot)$ is a nonlinear activation function, $\mathbf{W}^{\mathrm{int}}$ accounts for the reservoir connectivity, $\mathbf{W}^{\mathrm{inj}}$ defines the connectivity between the input and the reservoir, and $\mathbf{b}$ is a bias term that is usually randomly chosen for each reservoir node. Both, the matrix $\mathbf{W}^{\mathrm{int}}$ and $\mathbf{W}^{\mathrm{inj}}$ can be randomly chosen. In Eq. (\ref{eq:RC_output}), for the output layer, the output of the RC system $\mathbf{y}(t)$ is obtained as a linear combination of the reservoir states weighted with the output matrix $\mathbf{W}^{\mathrm{out}}$. The values of $\mathbf{W}^{\mathrm{out}}$ are obtained via a supervised learning procedure \cite{luko09}.

\subsection{Photonic RC leveraging spatio temporal dynamics} \label{sec:spatio}

\begin{figure*}[t]
	\centering
	\includegraphics[width=0.8\textwidth]{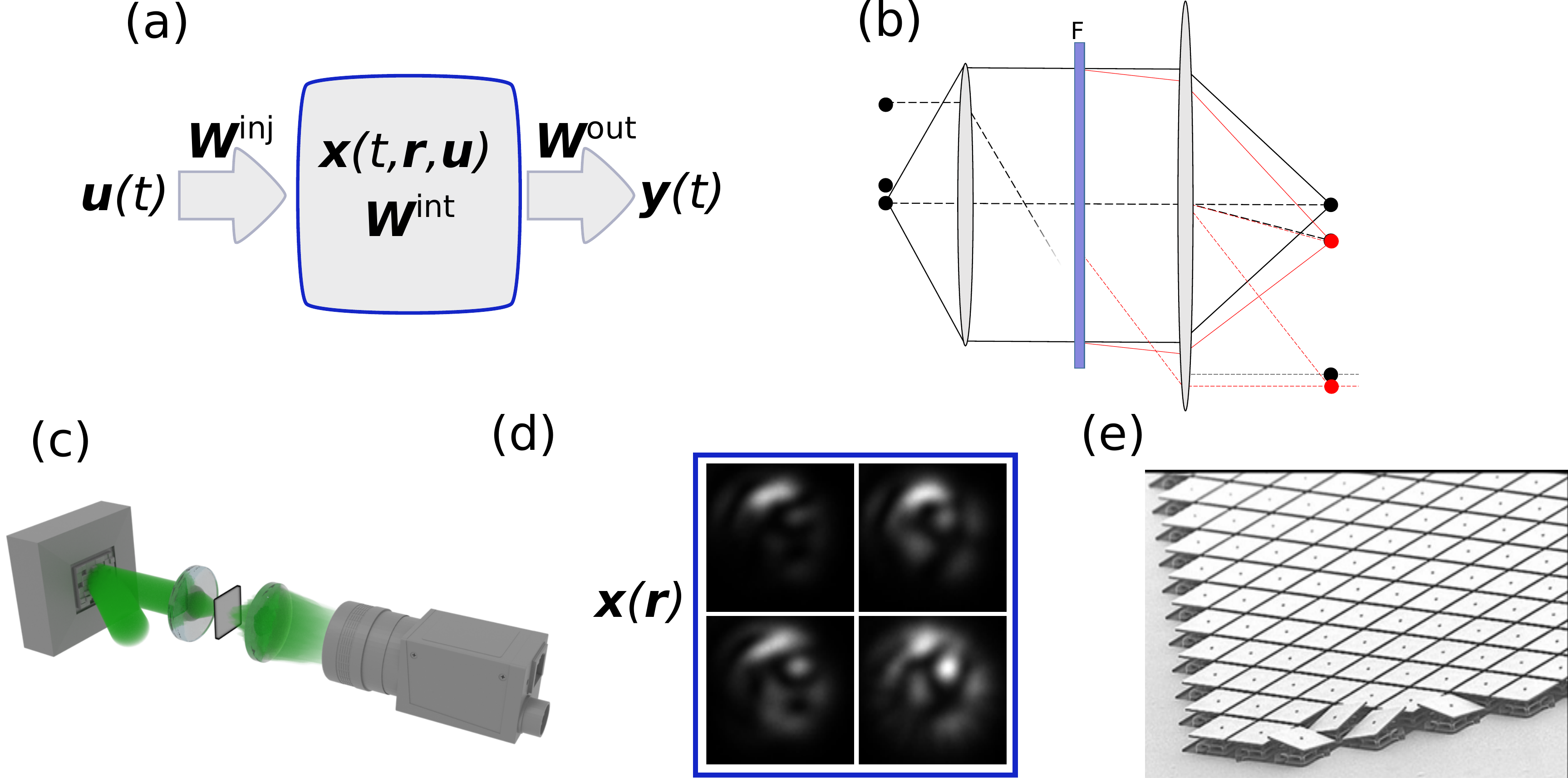}
	\caption{Spatio-temporal optical systems as RC.
		(a) Concept of spatially implementing a RC in a nonlinear medium.
		(b) Spatial parallelism in simple optical systems.
		(c) A spatial light modulator paired with a camera (from [18]) or (d) the complex spatial state of a multimode semiconductor laser (from [23]) can be used to implement a reservoir.
		(e) Finally, the output layer can readily implement parallel and programmable weights.}\label{fig:SpatioTempRC}
	%\hrule
\end{figure*}

The closest correspondence to the principle of RC according to Eqs. (\ref{eq:RC_state})-(\ref{eq:RC_output}) is to realize a driven network comprising nonlinear nodes whose state $\mathbf{x}(t)$ evolves for all dimensions in parallel.
This corresponds to all dimensions of $\mathbf{x}(t)$ to be physically represented simultaneously at each $t$, that they are simultaneously driven via input $\mathbf{u}(t)$, and that output $\mathbf{y}(t)$ can be created by the parallel physical realization of readout Eq. (\ref{eq:RC_output}).
This requires that the nonlinear dynamical system implements all connections $\mathbf{W}^{\mathrm{int}}$, $\mathbf{W}^{\mathrm{inj}}$ and $\mathbf{W}^{\mathrm{out}}$ simultaneously, see Fig. \ref{fig:SpatioTempRC}(a).
RC or NNs in general quickly approach $N=100\dots1000$ neurons or more, and in particular the number of internal connections scales with $O(N^2)$.
Massive parallelism to implement all connections is therefore essential.

Such parallelism can readily be achieved in optics based on spatial multiplexing, and the use of spatially distributed optical systems for communication and computing has been discussed starting in 1960 \cite{Cutrona1960,Weaver1966}.
A highly simple concept for such parallel connections leverages optical imaging, see Fig. \ref{fig:SpatioTempRC}(b).
Information spatially distributed at positions $\mathbf{r}$ in an object plane is transformed at the image plane.
The information capacity is determined through the system's diffraction limit ($\sim1~\mu\mathrm{m}^2$) and its field of view ($\sim10~\mathrm{mm}^2$), and hence 10$^8$ parallel dimensions are readily available for standard optical imaging arrangements \cite{Cutrona1960,Maktoobi2020}.
However, imaging alone only linearly maps a state $\mathbf{x}(t,\mathbf{r})$ onto a state $\mathbf{x}(t,\mathbf{r}')$.
Going beyond a linear map requires placing additional elements in the optical propagation path.

In the original proposition of linear optical computing, spatial filters realize optical convolutions fully in parallel \cite{Cutrona1960,Weaver1966}.
Following the same principle allows creating all internal RC connections, and the utilization for RC was first demonstrated based on diffractive coupling \cite{Brunner2015,Bueno2018} and later using complex optical media for implementing large scale random internal connections \cite{Dong2019,Rafayelyan2020}.
Integrated optical networks are also discussed and have been demonstrated.
Indeed, the first suggestion of photonic RC in 2008 was proposing integrated photonic networks of semiconductor optical amplifiers \cite{Vandoorne2008},  and a similar system was experimentally demonstrated in 2014 \cite{Vandoorne2014}.

One canonical approach creates the reservoir state in via iterative and repeated opto-electronic and electro-optical conversion based on a liquid-crystal spatial light modulator (SLM) and a camera.
Such a spatially distributed electro-optical RC was first demonstrated in 2018 \cite{Bueno2018} using diffractive orders \cite{Brunner2015,Maktoobi2020} to create a recurrent neural network featuring local coupling.
Denser connectivity was first demonstrated based on the same general experimental outline but using a thin layer of a complex optical scattering medium \cite{Dong2019}, see Fig. \ref{fig:SpatioTempRC}(c).
Recently, a multimode fiber operated at optical intensities sufficient for inducing nonlinear responses based on the Kerr effect implemented an extreme learning machine based on spatially distributed mode 
multiplexing \cite{Tegin2020}.

Optically multimode systems offer a convenient way to implement a reservoir's dimensions in spatial degrees of freedom.
Similar to the nonlinear multimode fiber based system \cite{Tegin2020}, recently a reservoir was implemented via discrete spatial sampling of a semiconductor large-area vertical-cavity surface-emitting laser (LA-VCSEL) \cite{Porte2021}, see Fig. \ref{fig:SpatioTempRC}(d).
Importantly, in this experimental demonstration, all conceptual sections of a photonic RC were implemented fully in parallel and real-time for the first time.
The complex transfer of a multimode fiber encoded the input weights $\mathbf{W}^{\mathrm{inj}}$, while a digital micro-mirror device implemented programmable readout weights $\mathbf{W}^{\mathrm{out}}$.
This experiment illustrates the benefit of spatially distributed NN implementations: increasing the NN's size will not reduce the computational speed, as all connections are realized fully in parallel.

\subsection{Delay-based Reservoir Computing} \label{sec:delay}

A practical scheme that enables RC with minimal hardware components was first introduced by Appeltant et al.~\cite{Appeltant2011a}. Here, a network of identical virtual nodes can be emulated out of a single physical nonlinear node and a long, linear delayed-feedback channel. The delay-based RC concept is very appealing for photonic implementations as it is straightforward to implement long delayed-feedback loops thanks to e.g. optical fibers. Using this delay-based RC concept, several hardware implementations could be accomplished using e.g. simple optoelectronic components~\cite{LAR12,paquot2012optoelectronic} or semiconductor lasers~\cite{brunner2013parallel}. These hardware implementations use off-the-shelf devices such that they can intrinsically benefit from robust, integrated, mature, and broadband photonic technologies that have been efficiently developed for Telecom applications during the last decades.

\begin{figure*}[thb]
	\begin{center}
		\includegraphics[width=0.6\textwidth]{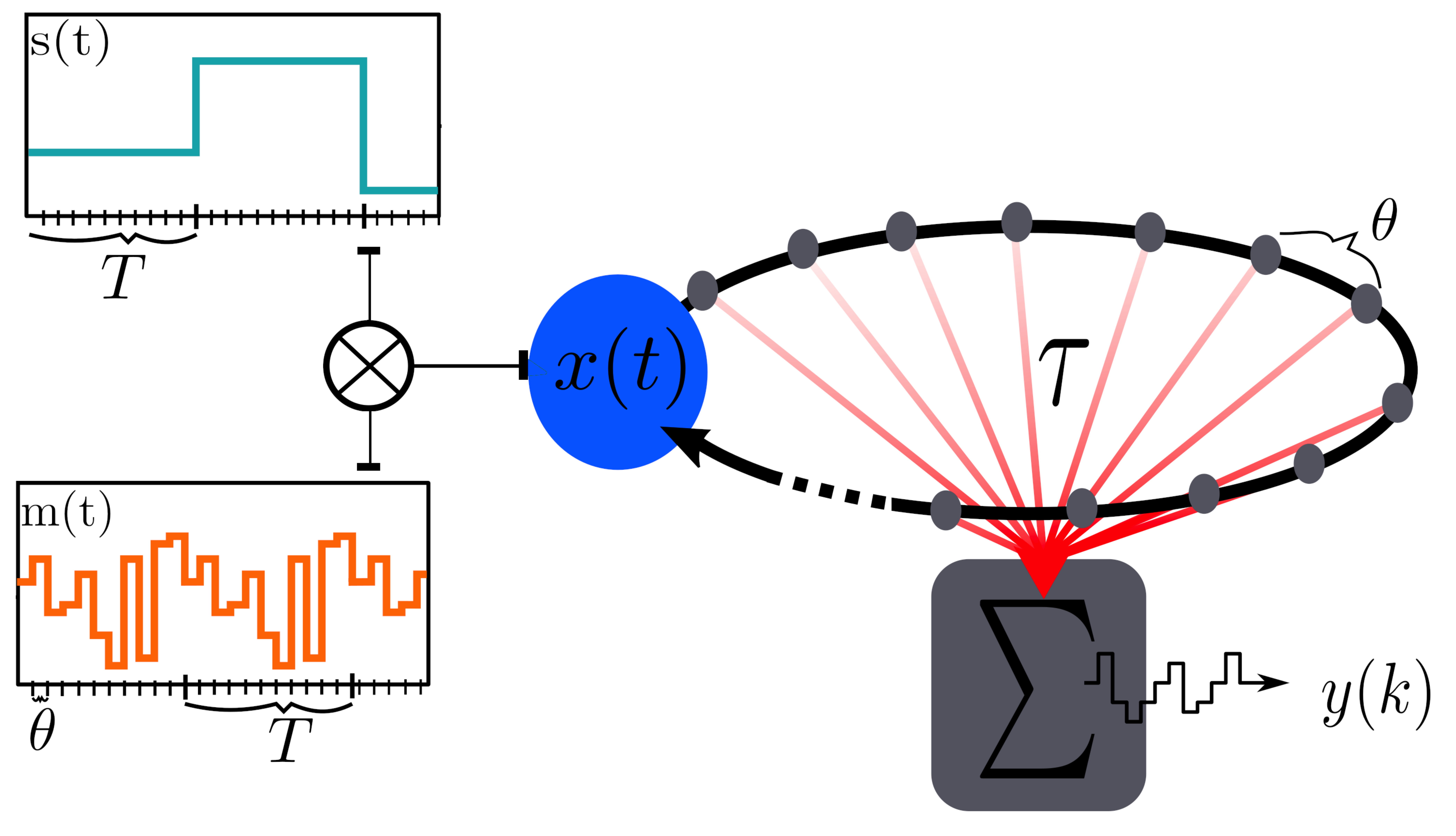}
	\end{center}
	\caption{Scheme of delay-based RC using a single nonlinear node with a delayed-feedback loop. The input signal $s(t)$ is time multiplexed using a mask $m(t)$. Using a node separation of $\theta$, a virtual network is created that contains $N=T/\theta$ virtual nodes. The states of the nodes are read out in the output layer and linear weights are trained to solve a given task. Figure courtesy of Mirko Goldmann.}
	\label{fig:delayRC}
\end{figure*}

For the delay-based scheme to work, the input data needs to be time-multiplexed. The reservoir network is then created in the temporal space. The basic principle used in delay-based RC schemes is derived from a very common concept in optical communication named time division multiplexing (TDM). It is, however, implemented under rather unusual dynamical conditions such that the so-called virtual nodes can be coupled with some of its neighbours. In TDM, different sources of information, originating from different places, are transmitted over a common line, e.g. on optical fiber channel, such that different information packets are allocated to different time slots. Important in TDM is to avoid any cross-talks between successive time slots, i.e. to prevent any information in one time slot (from one sender) from overlapping with the information from the previous or next time slot. This is achieved through a wide bandwidth of, both, the channel and the information processing system, which allows for preserving any unwanted mixing between neighbouring time slots. Delay-based RC, on the contrary, seeks mixing between temporal positions in order to couple the virtual nodes that are temporally emulated along a delay line. Such coupling is achieved either through a reduced bandwidth of the reservoir compared to the input information time scale \cite{Appeltant2011a}, or through de-syncrinization between feedback and information input temporal multiplexing \cite{paquot2012optoelectronic}.As illustrated in Fig. \ref{fig:delayRC},  the (virtual) nodes of the reservoir are taken at different temporal positions along the delayed-feedback loop. From a fundamental point of view, it is interesting that all the virtual nodes have identical dynamical properties as they all originate from the same nonlinear node, and that delay based reservoirs have a ring-like internal connection topology.

We describe here the procedure to process a temporal signal in the delay-based RC approach. For the sake of simplicity, we consider a time-discrete uni-dimensional input. In the input layer, each of the discrete inputs $s(k)$ is held for an input time $T$ to generate a continuous signal $\tilde{s}(t) = s(n)$ for $t \in[(k-1)T,kT]$, where $T$ is chosen close to the delay time $\tau$ in the feedback loop. The input signal $\tilde{s}(t)$ is subsequently multiplied by a step function $m(t)$ called input mask. The input mask is needed to enhance the diversity in the temporal responses of the nonlinear node. This mask often has a periodicity equal to the input time $T$ and it contains $N$ steps with a length $\theta$, where $N$ is also the number of virtual nodes. The so-called node separation $\theta$ determines the time the nonlinear node has to respond to the time-multiplexed input. If $\theta$ is chosen large compared to the intrinsic response time of the reservoir nodes, the latter response has enough time to settle down at a certain state \cite{paquot2012optoelectronic}. In contrast, for small node separation $\theta$, the response of each nonlinear node is always on a transient, each of them having a motion coupled to its close neighbours \cite{Appeltant2011a}. The amplitudes of the mask in between each $\theta$ are often drawn randomly from a uniform distribution, although chaotic \cite{nakayama2016laser} or other types of analog input masks \cite{argyris2021fast} have also been considered. The time-multiplexed input with the imprinted mask then drives the different virtual nonlinear nodes. In the output layer, the system is sampled with a frequency $1/\theta$ and the sampled states are used to train the reservoir's $N$ output weights $\mathbf{W}^{\mathrm{out}}$ using e.g. linear regression \cite{luko09}. Although the output layer is often implemented after digital post-processing, several attempts have been devoted to implement the output layer in analog hardware \cite{Antonik2017}.

\begin{figure*}[thb]
	\begin{center}
		\includegraphics[width=0.6\textwidth]{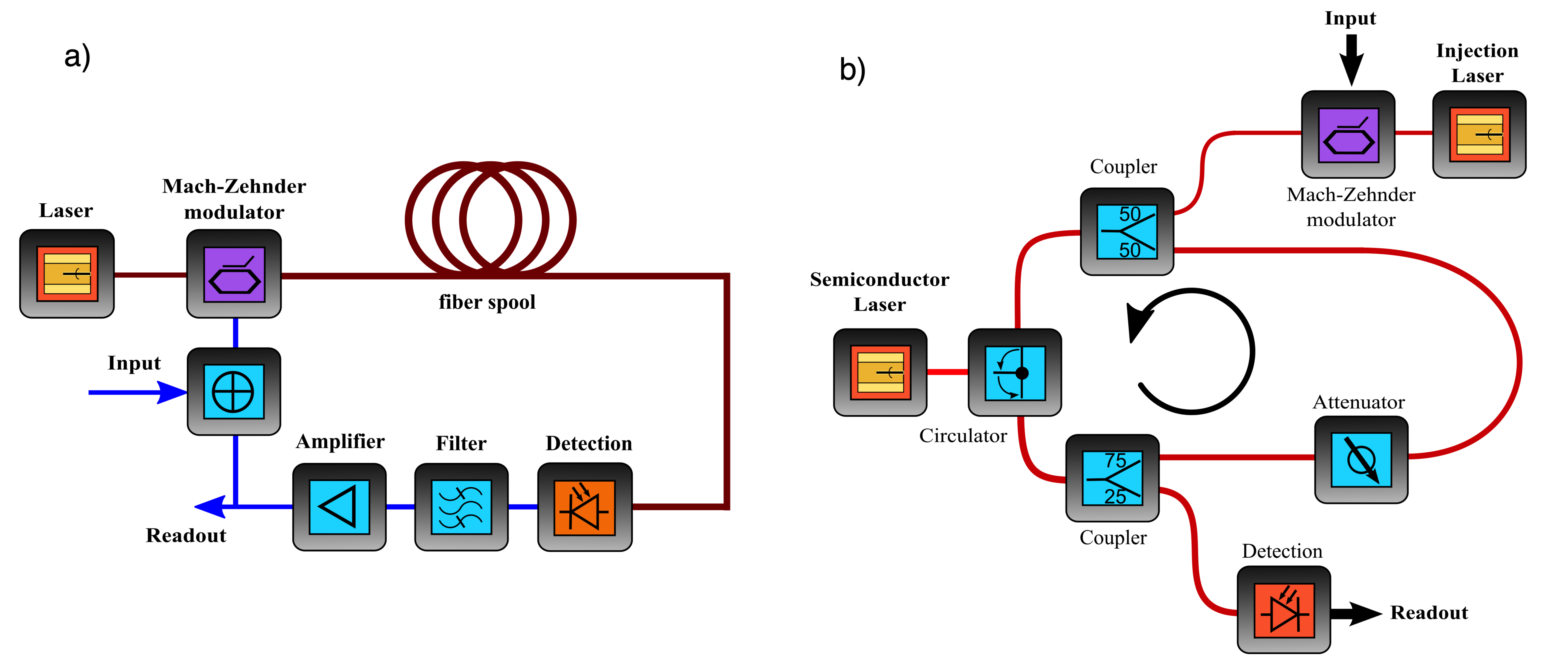}
	\end{center}
	\caption{a) Scheme of the optoelectronic RC based on a $sin^2$ nonlinearity. The optical (electronic) path is depicted in red (blue) color. The experimental setup comprises a laser diode, a Mach-Zehnder modulator as the reservoir node, a fiber spool for the delay, a photo diode for optoelectronic conversion, electronic filtering setting the nodes response time, and amplification defining the strength of the recurrence i.e. the feedback in the delay-based RNN. b) Scheme of photonic RC based on a semiconductor laser subject to delayed optical feedback. The experimental setup comprises the semiconductor laser as the reservoir node, a tunable laser source to optically inject the information, a Mach-Zehnder modulator, an optical attenuator, a circulator, couplers and a fast photo diode for signal detection.}
	\label{fig:schemes}
\end{figure*}

Figure \ref{fig:schemes} (a) shows a schematic representation of a delay-based optoelectronic implementation of RC. In this system, a Mach-Zehnder modulator provides an Ikeda-like $\sin^2$ nonlinearity and has been used e.g. to recognize spoken digits at a rate of 1 million words per second \cite{larger2017high}.
Figure \ref{fig:schemes} (b) illustrates a photonic RC implementation based on semiconductor lasers that have been used e.g. to decode coherent optical communication signals at high speeds \cite{estebanez2020accelerating}.

\section{Road map}

\subsection{Deep photonic neural networks}

In deep NNs \cite{LeCun2015} as well as deep RC \cite{Gallicchio2018}, the hierarchical arrangement of multiple nonlinear layers is of fundamental importance.
It provides the NN with chains of nonlinear transformations which are capable of highlighting features encoded in the input information that are encoded on different scales.
Such different input-data feature-scales can be encoded either be present in space or in time.

A first numerical investigation highlighted this importance for predicting chaotic systems by a cascade of nonlinear delay systems acting as the different layers of a deep RC \cite{Penkovsky2019}.
The intricate connection between computing properties such as the memory capacity and dynamical system aspects was investigated by Goldmann \emph{et al.} \cite{Goldmann2020}.
However, in deep RC, optimization is a challenge beyond simple tuning of hyper-parameters.
Other than in deep NN's optimized on back-propagation of error gradients, the connections between the individual layers of a deep RC cannot be optimized straight away as the optimization of $\mathbf{W}^{\mathrm{out}}$ does not provide gradients.
This challenge was addressed by Freiberger \emph{at al.} \cite{Freiberger2020}, where the authors optimized intra-layer connections offline on a standard computer via back-propagation and then investigated how these weights would perform when transfered to hardware.
Such optimized weights do improve a deep RC's performance, however, manufacturing tolerances inherent to hardware implementations can significantly reduce the beneficial nature of such pre-training.

Finally, recently a spatio-temporal electro-optical NN layer demonstrated fully hardware implemented training and inference of a deep NN \cite{Zhou2021}.
In their work, the author's used a DMD to encode input information.
The collimated reflected signal was then phase-modulated via an SLM before detection with a camera.
For training, the SLM's phase mask was iteratively updated by merging beam-propagation calculations and error back-propagation.
Using their setup, the authors emulated a variety of deep NN architectures by using the recorded camera image of a preceding layer as the input for the DMD of the consecutive layer.
Importantly, their concept's performance approached a fully numerical implementation of the famous LeNet-5, and was superior in computing capacity and energy efficiency to a state of the art graphical processing unit \cite{Brunner2021}.

\subsection{Training photonic neural networks}

Most approaches for training photonic neural networks rely on a supervised learning procedure, i.e. the system learns from examples. On the one hand, deep neural networks are usually trained via backpropagation of errors, and recurrent neural networks often via backpropagation through time. These training methods are routinely implemented in software but it is challenging to migrate these methods to photonic hardware.

In particular, RC training methodology is more hardware-friendly as only the output weights are modified in the learning process. Most photonic implementations of the RC concept resource to off-line digital post-processing to obtain the optimum values of the output weights. There are several attempts to go around this off-line training by e.g. using a simple gradient descent algorithm to optimize the output weights online \cite{antonik2016online} or using a reinforcement learning algorithm to approximate the desired outputs with Boolean output weights \cite{Bueno2018}.

Going beyond a training restricted to the output layer, an interesting possibility to train the connections between all layers of the photonic neural networks is to rely on accurate models of the physical systems. In this manner, the optimum values of the connection weights can be transferred to the physical system after the training has been done in the numerical counterpart. This idea was tested on an optoelectronic version of a delay-based reservoir computer \cite{hermans2015photonic}, where there is a good agreement between the ideal model and the physical implementation. In \cite{hermans2015photonic}, the performance of the photonic system was indeed more accurate when fully trained than when only the output layer was trained. 

A crucial improvement for training photonic neural networks consists in realizing the backpropagation algorithm by physical means. This concept was originally presented in \cite{wagner1987multilayer} for multilayer optical neural networks, where a physical error signal that propagates from output to input can be measured to update the system parameters. More recently, a similar concept has been suggested for all-optical training of neural networks using saturable absorption for the nonlinear units \cite{guo2021backpropagation}. In this context, the possibility to have backpropagation training methods that are hardware-friendly is becoming a reality \cite{wright2021deep,lopez2021self}.
For the training of photonic neural networks, bio-inspired learning methods like equilibrium propagation \cite{scellier2017equilibrium} or direct feedback alignment \cite{Launay2020} are interesting and promising options.

\subsection{Autonomous photonic neural networks}

Finally, the fundamental appeal of photonic NNs lies in boosting energy efficiency, speed and reducing latency by realizing connections between the artificial neurons optically.
As in modern optical communications, the countless connections between the neurons of a NN of competitive size correspond to a massive communication network, and photonic signal transduction has fundamental advantages compared to electronic communication \cite{Lohmann1990}.
Exploiting these fundamental advantages of photonics for NNs therefore ultimately requires that digital electronic computers are not significantly involved in computing a neuron's input state.

The input of a neuron corresponds to multiplying the state of the preceding neuron or input channel with the connection's weight, i.e. a multiplication.
The result is then to be added to the product of the next weighted input channel of the same neuron, and one therefore speaks of multiply and accumulate (MAC) as the fundamental unit of computing operations in NNs.
Photonic NNs in which a digital electronic computer is required at some stage during each MAC limits bandwidth and parallelism to the one of the auxiliary computer.
The required electronic computer therefore needs to run at very high realtime bandwidths to exploit the high photonic signal rates, and for spatio-temporal systems the computer would have to measure and store numerous values for each artificial neuron.
The step from a required high-performance electronic computer only acting as auxiliary system for the photonic NN towards emulating the entire NN fully on this electronic substrate quickly becomes negligible.
In such conditions, the beneficial contribution of the photonic section in terms of bandwidth and energy efficiency usually is marginalized, or even becomes counter productive.

Future photonic NNs will always work in collaboration with electronic circuitry.
However, it is crucial that ultimately the active involvement of auxiliary electronic hardware is moved away from the single neuron response level.
Electronic hardware will most likely be required on the level of the photonic NN's input and output information, the general management of the photonic hardware and most likely for implementing learning concepts.

\section{Outlook}

Currently, photonic systems are being discussed as accelerators for information processing in electronic computing \cite{kitayama2019novel}.
For the future success of nonlinear photonics as main drivers for unconventional computing implementations, several challenges remain ahead.
Aspects of size scalability, speed and power consumption will play a major role in the ultimate prevalence of photonics in the age of big data computing.

\section*{Acknowledgments}

We acknowledge the Spanish State Research Agency, through the Severo Ochoa and Mar\'ia de Maeztu Program for Centers and Units of Excellence in R\&D (MDM-2017-0711) and through the QUARESC project (PID2019-109094GB-C22/ AEI / 10.13039/501100011033), and through the Region Bourgogne Franche-Comté, the French National Research Agency via EUR EIPHI program (Contract No. ANR-17-EURE- 0002), and the Volkswagen Foundation (NeuroQNet II) and the European Union’s Horizon 2020 research and innovation program under the Marie Skłodowska-Curie Grant Agreements No. 860830 (POST DIGITAL). The work of MCS has been supported by MICINN/AEI/FEDER and the University of the Balearic Islands through a ``Ramon y Cajal'' Fellowship (RYC-2015-18140).

\bibliography{references}
\bibliographystyle{ieeetr}

\end{document}